# Observational evidence of magnetic reconnection in the terrestrial bow shock transition region

Shan Wang[1,2], Li-Jen Chen[2], Naoki Bessho[1,2], Michael Hesse[3,4], Lynn B. Wilson III[2], Barbara Giles[2], Thomas E. Moore[2], Christopher T. Russell[5], Roy B. Torbert[6], and James L. Burch[4]

[1]Astronomy Department, University of Maryland, College Park, MD 20742

[2]NASA Goddard Space Flight Center, Greenbelt, MD 20771

[3]Department of Physics and Technology, University of Bergen, Bergen, Norway

[4]Southwest Research Institute San Antonio, San Antonio, TX 78238

[5]Department of Earth, Planetary, and Space Sciences, University of California, Los Angeles, CA 90095

[6]Space Science Center, University of New Hampshire, Durham, NH 03824

*swang90@umd.edu

Key points

- Magnetic reconnection is observed in the shock transition region
- Features of a reconnection diffusion region are observed: electron outflow jet, Hall flow and fields, and enhanced energy conversion rates
- Ion energization to suprathermal energies and parallel electron heating are found to be consistent with a reconnection exhaust

Abstract

We report evidence of magnetic reconnection in the transition region of Earth's bow shock when the angle between the shock normal and the immediate upstream magnetic field is 65 degrees. An ion-skin-depth-scale current sheet exhibits the Hall current and field pattern, electron outflow jet, and enhanced energy conversion rate through the nonideal electric field, all consistent with a reconnection diffusion region close to the X-line. In the diffusion region, electrons are modulated by electromagnetic waves. An ion exhaust with energized field-aligned ions and electron parallel heating are observed in the same shock transition region. The energized ions are more separated from the inflowing ions in velocity above the current sheet than below, possibly due to the shear flow between the two inflow regions. The observation suggests that magnetic reconnection may contribute to shock energy dissipation.

1. Introduction

Collisionless magnetic reconnection and shocks are two classes of phenomena in which energy dissipation and plasma energization are fundamentally important. Global kinetic simulations have shown that at quasi-parallel shocks, where the angle ($\theta_{Bn}$) between the upstream magnetic field and the shock normal is smaller than 45°, shock reflected ions can propagate upstream, interact with incident solar wind ions, and become unstable (e.g., Karimabadi et al., 2014; Wilson III, 2016). The foreshock region with the existence of back-streaming ions, the shock transition region close to the main ramp, and the downstream of the shock thus become 'turbulent' and contain many current structures. In a shock hybrid simulation with parameters relevant to the Earth's quasi-parallel shock

(Alfvén Mach number $M_A$=8), reconnection is shown to occur in the shock transition region (Gingell et al., 2017). Reconnection is also demonstrated in particle-in-cell simulations relevant to astrophysical (high $M_A$~40) quasi-perpendicular shocks ($\theta_{Bn}$>45°), where current sheets are generated through the Weibel instability, and the magnetic islands generated by reconnection accelerate electrons to suprathermal energies (e.g., Matsumoto et al., 2015). Evidence of reconnection has been reported to occur in the magnetosheath downstream of the Earth's bow shock (Retinò et al., 2007; Vörös et al., 2017; Phan et al., 2018; Wilder et al., 2017, 2018; Eriksson et al., 2018). Features of current sheet structures consistent with magnetic reconnection were identified in the transition region of a quasi-parallel shock in a recent study (Gingell et al., 2018). The role of reconnection in energy conversion in the shock, and differences and similarities between the reconnection features at the shock and those in magnetotail and magnetopause are yet to be investigated. Shock parameters (e.g., $\theta_{Bn}$ and $M_A$) affect the properties of particle distributions, electromagnetic waves and instabilities (e.g., Krasnoselskikh et al., 2013; Burgess and Scholer, 2013), and may also make a difference on whether and how reconnection occurs in the shock.

In this study, we show evidence of reconnection in current sheets residing in the bow shock transition region before the plasma flow is fully decelerated to the downstream level. The shock has an immediate upstream $\theta_{Bn}$ of approximately 65 degrees. Features of the reconnection diffusion region and the ion exhaust are discovered to be consistent with those in symmetric reconnection.

2. Data

The observation data are from the Magnetospheric Multiscale mission (MMS) (Burch et al., 2016). Plasma data are from the Fast Plasma Investigation instrument (Pollock et al., 2016), with 150 ms resolution for ions and 30 ms resolution for electrons. Magnetic fields for the wave analysis (Figure 2i) are from the Search Coil Magnetometer (SCM) with 8192 samples/s (Le Contel et al., 2016). DC Magnetic field measurements are from the FluxGate Magnetometer (FGM) (Russell et al., 2016) with 128 samples/s in the burst mode. Electric field data are from the Axial (Ergun et al., 2016) and Spin-plane double probes (Lindqvist et al., 2016) with 8192 samples/s.

3. Observations

3.1 Overview of the shock event

MMS crossed the bow shock at GSE (8.4, 8.4, 0.1) $R_E$ in the dusk sector around 13:30 UT on 9 November, 2016 (Figure 1f). Taking the interval around 13:20-13:21 UT when backstreaming ions (above a few keV, Figure 1a) are absent as the upstream region, the angle ($\theta_{Bn} = \arccos(|\boldsymbol{b}_u \cdot \boldsymbol{n}|)$) between the upstream magnetic field direction ($\boldsymbol{b}_u$) and the bow shock normal ($\boldsymbol{n}$) based on *Farris et al.* (1991) model is 65°. Using data from the same upstream interval and from the downstream magnetosheath during 13:29:10-13:29:40 UT, where the ion bulk speed has reduced to a quasi-steady level of ~110 km/s, $\theta_{Bn}$ is obtained to be 57°-60° based on two other methods: (1) the iterative least square technique based on the Rankine-Hugoniot conservation equations (Koval et al., 2008); (2) the co-planarity of the shock normal and the jump of upstream and downstream $\boldsymbol{B}$ and $\boldsymbol{V}$ (Schwartz,

2000). The upstream Alfvén Mach number is 11, and the magnetosonic Mach number is 4.

A turbulent foreshock region exists at, e.g., 13:21-13:26 UT. Upon entering the foreshock ~13:21 UT, the incoming solar wind population has $v_{//}>0$ along the magnetic field (red population in the example distribution in Figure 1g); suprathermal ions appear with $v_{//}<0$ (blue population), indicating that they are moving towards upstream. Magnetic fields exhibit significant fluctuations. In particular, the magnetic field and density pulses around 13:23-13:26 UT are consistent with the short large amplitude magnetic structures (SLAMS, e.g., Schwartz et al., 1992; Mann et al., 1994; Wilson III et al., 2013). Such a turbulent foreshock with transient events is commonly observed at quasi-parallel shocks with $\theta_{Bn}<45°$ (e.g., Eastwood et al., 2005; Wilson III, 2016). MMS was in the solar wind for about 20 min before 13:20 UT and observed variations of magnetic fields that occasionally make $\theta_{Bn}$ approach $45°$ (not shown). Even though the immediate upstream condition measured by MMS in the dusk sector is quasi-perpendicular, we note the possibility that the shock geometry may vary toward quasi-parallel after 1320 UT.

As will be analyzed below (Figures 2-3), the two current sheets between the dashed vertical lines show evidence consistent with reconnection. During the marked interval, the bulk $V_{ix}$ (Figure 1c) is about -150 km/s, not yet reaching the downstream value of about -60 km/s near 13:30 UT, while the density (Figure 1b) and the magnetic field strength (Figure 1d) are about 5 and 8 times of the upstream values, respectively. Thus, the location is interpreted to be within the transition region before reaching the magnetosheath proper.

## 3.2 Reconnection diffusion region encounter

The zoom-in plot (Figure 2) shows that current sheet A exhibits features consistent with a reconnection diffusion region. The electron velocity, magnetic field, and electric field in Figures 2b-2d are in the LMN coordinate determined by the Minimum Variance Analysis (MVA) of magnetic fields during 13:39:25.69-13:39:26.42 UT. $V_{eL}$ has a negative peak (location 1) close to the $B_L$ reversal with an amplitude of about 150 km/s~$1.5V_A$ larger than the ambient value, supporting the interpretation of a reconnection outflow jet. Here the ion Alfvén speed ($V_A$) based on $B_L$ of 40~50 nT and $n$ of 100 cm$^{-3}$ is about 100 km/s. In other words, in the $B_L>0$ region with a yellow shade, $V_{eL}$ decreases and $B_L$ decreases (positive correlation); in the $B_L<0$ region with a light blue shade, $V_{eL}$ increases and $B_L$ decreases (negative correlation). Such a change of the $V_L$-$B_L$ correlation in a current sheet is commonly used to identify reconnection outflow jets (Gosling et al., 2005; Phan et al., 2018). Near the $B_L$ reversal, $B_M$ is vanishingly small, suggesting a negligible guide field. Based on the four-spacecraft timing analysis at the $B_L$ reversal, the half-thickness of current sheet A (in the interval marked by the green bar in Figure 2c) is estimated to be 1 $d_i$ or 0.8 $r_{gi}$, where the ion skin depth ($d_i$) is ~23 km ($n$=100 cm$^{-3}$) and ion thermal gyroradius ($r_{gi}$) is ~28 km (for $T_\perp = 130\ eV$ and $|\mathbf{B}|$=40 nT).

The bipolar $B_M$ and bipolar $E_N$ are consistent with the Hall electric and magnetic fields expected in symmetric reconnection (e.g., Drake et al., 2008), illustrated in Figure 2m. Concurrent with the Hall fields, close to 13:39:25.9 and 13:39:26.2 UT, $V_{eL}$ is less negative than ambient values, supporting an X-line directing flow pattern sandwiching the outflow jet. The X-line directing electron flows and the outflow together form the Hall current loop as in magnetotail reconnection (e.g., Nagai et al., 2001).

The local energy conversion through the non-ideal electric field $J \cdot E' = J \cdot (E + V_e \times B)$ (Figure 2e) is enhanced mainly in the perpendicular component near the mid-plane (location 1), a feature consistent with laboratory (Yamada et al., 2016), space (Wilder et al., 2018), and simulation (Yamada et al., 2016) results of energy conversion in the diffusion region of reconnection with symmetric upstream conditions and negligible guide field. The enhancement of $J \cdot E'$ occurs in the parallel component near the separatrix in the Hall structure (location 2). In summary for current sheet A, the electron flow pattern, the Hall fields, and the enhancement of energy conversion through $E'$ are consistent with a reconnection diffusion region.

Throughout the current sheet crossing, parallel and perpendicular temperatures oscillate 180 degrees out of phase with each other at a frequency of ~5 Hz, while the total temperature remains roughly constant (Figure 2g). Electron velocity distribution functions (VDFs) at locations 1 and 2 with $J \cdot E'$ enhancements are shown in Figures 2j-2k. VDF 1 is mostly isotropic, and a slight asymmetry along $v_{\perp 1}$ leads to a perpendicular flow (marked by the horizontal dotted line in Fig. 2h). VDF 2 is elongated along $v_{//}$. The electron pitch angle (PA) distribution for 25-100 eV (3000-6000 km/s) electrons (Figure 2h) further demonstrates that $T_{e//}$ peaks correspond to enhancements of parallel and anti-parallel phase-space densities (psd) similar to VDF 2; $T_{e\perp}$ peaks correspond to enhancements at PA~90°, where VDFs become quasi-isotropic.

The $T_e$ oscillation is associated with electromagnetic waves. Electric and magnetic fields during the $T_e$ oscillation exhibit broadband power spectra below the upstream lower-hybrid frequency ($f_{lh}$~30 Hz) (not shown). The bandpass magnetic fields over 4-6 Hz in the field-aligned coordinates (Figure 2i) show that the fluctuation along the background magnetic field $B_0$ ($\delta B_\parallel$, red) is mostly anti-correlated with the density fluctuation ($\delta n$, black). The minima of $\delta B_\parallel$ are collocated with the PA~90° psd enhancements and quasi-isotropic VDFs (Figure 2h), supporting the modulation of electrons by the waves. The observed anti-correlation between $\delta B_\parallel$ and $\delta n$ and wave properties such as the ion-frame frequency and phase velocity indicate consistencies with the kinetic Alfvén wave (KAW). Details about the wave property analyses are provided as supplementary information. The waves only modulate $T_e$ components without net heating, which suggests a limited role of such waves in the net energy conversion during reconnection.

Figure 2b shows that the ion velocity (dashed curves) does not differ in any significant way from the ambient value. Concurrently with local $\boldsymbol{J} \cdot \boldsymbol{E}'$ maximum (location 1), a local peak in $T_{i\perp}$ and a dip in $T_{i/\!/}$ occur, while $T_{it}$ only has a small enhancement (~4.5 eV) compared to the previous point (Figure 2f). The variation of the $T_i$ components is largely due to the magnetic field direction changes, while the VDF in LMN does not alter much over time (not shown). The ion VDF at location 1 in the $V_M$-$V_N$ plane (Figure 2l) shows an energetic tail in addition to the core distribution. The typical diffusion region VDF features, like elongation along $M$ due to acceleration by the reconnection electric field, or counter-streaming $v_N$ due to in-plane potential (e.g., Wang et al., 2016a), are not clearly present, though the core part of the distribution exhibits some distortion. It is not clear yet whether ions do not respond to such a reconnection structure like suggested in Phan et al. (2018), or the energization feature cannot be clearly observed due to high $\beta_i=2$ upstream of the current sheet.

3.3 Reconnection exhaust encounter

Energized ions moving along the magnetic field and $\boldsymbol{E} \times \boldsymbol{B}$ drifting with the incoming solar wind are observed for an extended interval including current sheet B (Figure 3). These energized ions are consistent with a reconnection exhaust when the outflowing ions have been remagnetized. The velocity, magnetic and electric fields (Figure 3c-3f) are transformed to an LMN coordinate using MVA during 13:39:18.21-13:39:24.74 UT. The guide field is estimated to be about 20%, based on the magnetic field at the start and end of the interval ($B_M$~10 nT, $B_L$~50 nT). The half-thickness for the interval marked by the green bar in Figure 3e is about 4 $d_i$. As shown in the $v_{iL}$ spectrogram (Figure 3a) and the bulk $V_{iL}$ (dashed curves in Figure 3c), the background flows outside of the current sheet have a sub-Alfvénic difference between the $B_L<0$ side (~13:39:19 UT) and the $B_L>0$ side (~13:39:25 UT). The average flow ([-195, 52, -55]km/s in LMN) between the two sides is taken to be the X-line motion, as expected for symmetric upstream density and magnetic field conditions (Doss et al., 2015). The velocity (Figure 3d) after subtracting the X-line motion exhibits enhancements towards negative $V_L$ by about -50 km/s (~0.5 $V_A$) for both ions and electrons near the $B_L$ reversal, suggesting a reconnection outflow jet at the -L side of the X-line (illustrated in Figure 3o with an orange arrow). The light blue and yellow shaded regions exhibit negative and positive $V_{iL}$-$B_L$ correlations, respectively. The increase of $E_N$ (Figure 3f) near the current sheet mid-plane is associated with the enhanced $V_L$ and guide field $B_M$. The ion and electron bulk velocities only differ in the $M$ direction by about 80 km/s (~$0.8V_A$), leading to a current density parallel to the magnetic field at the mid-plane. The local energy conversion $\boldsymbol{J} \cdot \boldsymbol{E}'$ is negligible (Figure 3g).

The $v_{iL}$ spectrogram at the $B_L>0$ side exhibits two populations. The VDF at location 5 (Figure 3m) shows that the two populations are streaming along $v_{//}$ and follow the same $\boldsymbol{E} \times \boldsymbol{B}$ drift. Population $i$ (indicating 'inflowing' component) has a bulk $V_{//}$ of -120 km/s and a similar intensity as the solar wind ions outside of the current sheet; population $o$ (symbol for 'outflow' away from the X-line) has a bulk $V_{//}$ of -360 km/s. Population $o$ travels towards the Earth. For shock reflected ions, if magnetized, they are expected to follow the $\boldsymbol{E} \times \boldsymbol{B}$ drift and the parallel velocity is towards the upstream (so-called 'field-aligned beams' (e.g., Fuselier, 1995)); if demagnetized, they may have a bulk velocity towards downstream when gyrating around the magnetic field, appearing to be non-gyrotropic in VDFs (e.g., Sckopke et al., 1990). Therefore, population $o$ that follows the $\boldsymbol{E} \times \boldsymbol{B}$ drift and moves Earthward is not likely due to shock reflection, and is interpreted as due to reconnection. The mixing of the two field-aligned populations $\boldsymbol{E} \times \boldsymbol{B}$ drifting together is interpreted as a consequence of reconnection as it resembles distributions predicted by theories and simulations (Drake et al., 2009) and observations (e.g., Hietala et al., 2015; Wang et al., 2016b). The co-existence of the two populations with a relative drift along $v_{//}$ leads to an increase of the parallel ion temperature (Figure 3h).

The ion and electron VDF evolution at the $B_L>0$ side further substantiates the interpretation of the reconnection exhaust. A deHoffmann-Teller (HT) velocity is determined using velocities and magnetic fields for the shown interval to be $V_{HT}$=[-249, 41, -57] km/s in LMN. In the HT frame, where the perpendicular motional electric field is transformed away and the parallel electric field pointing away from the X-line (illustrated in Figure 3o with gray arrows) accelerates (decelerates) electrons (ions) towards the mid-plane (Haggerty et al., 2015). In ion VDFs, the white vertical dashed lines mark the parallel component of the deHoffmann-Teller velocity ($V_{HT//}$). For VDF 5 in the HT frame, population $i$ has $v_{//}>0$ (centered at 110 km/s) moving towards the X-line, and population $o$ has $v_{//}<0$ (centered at -130 km/s) away from the X-line (illustrated in Figure 3o). In VDF 4' (closest ion measurement to $B_L$=0), the centroid of the most intense population has $v_{//}$=60 km/s in the HT frame, with a parallel energy reduction of 44 eV compared to VDF 5. On the other hand, the electron VDF along $v_{//}>0$ is more elongated at location 4 (Figure 3n, black) than at location 5 (Figure 3n, red), and the difference can be accounted for by a 40 eV potential in the spacecraft frame (~38 eV in HT frame) as demonstrated by the green curve. The value of the parallel potential estimated from ion and electron VDFs approximately agree with each other.

The ions above and below the current sheet differ significantly in their $v_{//}$ distributions, as seen from the $v_{//}$ spectrogram in the HT frame (Figure 3j). The difference may be in part due to the shear flow between the two inflow regions. In the HT frame illustrated in Figure 3o, the inflow in $B_L>0$ has a larger $V_L>0$ and $V_{//}>0$, and the inflow in $B_L<0$ has $V_L>0$ and $V_{//}<0$ with smaller amplitudes. In the exhaust, inflowing ions (population $i$ from $B_L>0$ and population $i'$ from $B_L<0$, labeled in Figure 3j with red letters) are gradually decelerated towards the mid-plane vicinity. Because of the small initial $|V_\parallel|$, the core part of population $i'$ may be decelerated to $V_\parallel=0$ and turned to $V_\parallel>0$ before reaching the mid-plane (e.g., ~13:39:20 UT), while the core part of population $i$ with a large $V_\parallel$ may penetrate to the $B_L<0$ side. VDFs near location 3 (Figure 3k) may contain a mixture of populations $i$ and $i'$. The fact that the two populations are not well separated also support that $E_\parallel$ in $B_L<0$ is not as significant as in $B_L>0$. The high-energy part of population $i'$ may reach the mid-plane and $B_L>0$ regions, contributing to population $o$.

We also note that the guide field effect cannot explain the observed asymmetry. A guide field along the out-of-plane current direction causes $E_\parallel$ at the $-L$ side of the X-line mostly positive, which would accelerate ions (electrons) towards the $B_L<0$ ($B_L>0$) side (e.g., Le et al., 2009; Eastwood et al., 2018, and references therein), inconsistent with the observation (e.g., ions are accelerated from the mid-plane towards $B_L>0$).

4. Summary and discussions

In this study, we report evidence of two reconnecting current sheets in the transition region of a bow shock. Current sheet A is consistent with the diffusion region in symmetric reconnection with a negligible guide field, showing features: 1. an electron outflow jet and the Hall flow pattern, 2. the bipolar Hall magnetic and electric fields across the current sheet, 3. enhanced energy conversion through the nonideal electric field. The half-width of current sheet A is ~ 1 $d_i$. Energized field-aligned ions consistent with a reconnection exhaust are observed in association with current sheet B, which exhibits also parallel electron heating and has a thickness of a few $d_i$. How these reconnecting current sheets are generated is an important question. Theory suggests that at high $M_A$ (>10) quasi-perpendicular shocks, reconnecting current sheets can be generated through nonlinear saturation of the Weibel instability (Matsumoto et al., 2015). A hybrid simulation shows that reconnection in the transition region of a marginally quasi-parallel shock ($\theta_{Bn}$=40°, $M_A$=8) is born out of the turbulent magnetic field fluctuations associated with back-streaming ions (Gingell et al., 2017). The reported shock with $M_A$~11 marginally satisfies the criterion of the Weibel instability generating reconnection, and exhibits turbulent magnetic field structures associated with back-streaming ions. Further investigations are required to reveal the generation mechanisms of reconnecting current sheets and the dependence on shock parameters.

Below we attempt to discuss the role of reconnection in the energy conversion for the observed event. For the shock where the flow energy is the dominant form of the upstream energy, the available energy per pair of plasmas is $\frac{1}{2}m_i V_n^2$ (e.g., Schwartz et al., 1987), where $V_n$ is the normal component of the upstream velocity, which is 470 eV for the presented shock. For reconnection, the available electromagnetic energy per particle is $m_i V_A^2$ in the inflow region, and the fraction of thermal energy gain can be approximated to be $\frac{\gamma}{\gamma-1}\frac{k_B T}{m_i V_A^2}$ (e.g., Phan et al., 2013, 2014; Shay et al., 2014; Wang et al., 2018), where $\gamma$ is the adiabatic index. For the shown reconnecting current sheet B, $m_i V_A^2$ is 104 eV. For current sheet B, considering the interval of 13:39:19-13:39:25 UT with ion mixture as the reconnection exhaust, the average ion temperature in the exhaust is increased by 12.6 eV compared to the average inflow temperature outside of the exhaust boundaries, and electron temperature increase calculated in the analogous way is 1.2 eV. Taking $\gamma$ to be 5/3, the fraction of thermal energy gain for such a reconnection structure is 33%. It suggests that for the involved particles, reconnection contributes to converting 104/470*33%~ 7% of the upstream energy to the thermal energy. As discussed in section 3, no clear net energization is observed in current sheet A. Thus, statistical studies of energy conversion properties in reconnecting current sheets in the shock transition region is needed.

In order to further address the role of reconnection in the overall shock dissipation, we need to consider the occurrence rate of reconnection in the shock structure. The presented two reconnecting current sheets are identified mainly based on the signature of outflow jet and the Hall features. The outflow jet for symmetric reconnection is peaked at the current sheet mid-plane, such that $V_L$ and $B_L$ have positive and negative correlations on the two sides of the mid-plane, respectively. Among the ~50 current sheets in the transition region of the presented shock event, only the presented current sheets reliably satisfy such *V-B* correlation criterion. On the other hand, the electron flow pattern in asymmetric reconnection has been shown by both PIC simulations and MMS measurements (e.g. Chen et al., 2016) to distinguish from that in the symmetric case in that the flow in the separatrix region is toward the X line on the high density side and away from the X-line in the low density side. Outside of the diffusion region, the current sheet can be mostly viewed as one rotational discontinuity like at the magnetopause, where the MHD flow tangential to the current sheet and magnetic field maintain a single correlation (e.g., Sonnerup et al., 1981). Therefore, both inside and outside the diffusion region, current sheets undergoing asymmetric reconnection will not exhibit the sequential correlation and anti-correlation between *V* and *B*. In the shock transition region, since the plasma flux mainly comes from the incoming solar wind towards Earth, reconnection may be asymmetric. Current sheets not exhibiting the *V-B* correlation-anticorrelation at shocks may still be reconnecting. Further investigations with the aid of simulations are required to better characterize reconnecting current sheets at the shock transition region.

Acknowledgments

The research is supported in part by a DOE grant DESC0016278, NSF grants AGS-1619584, AGS-1552142, a NASA grant 80NSSC18K1369 and the NASA MMS mission. MMS data are available at MMS Science Data Center (https://lasp.colorado.edu/mms/sdc/).

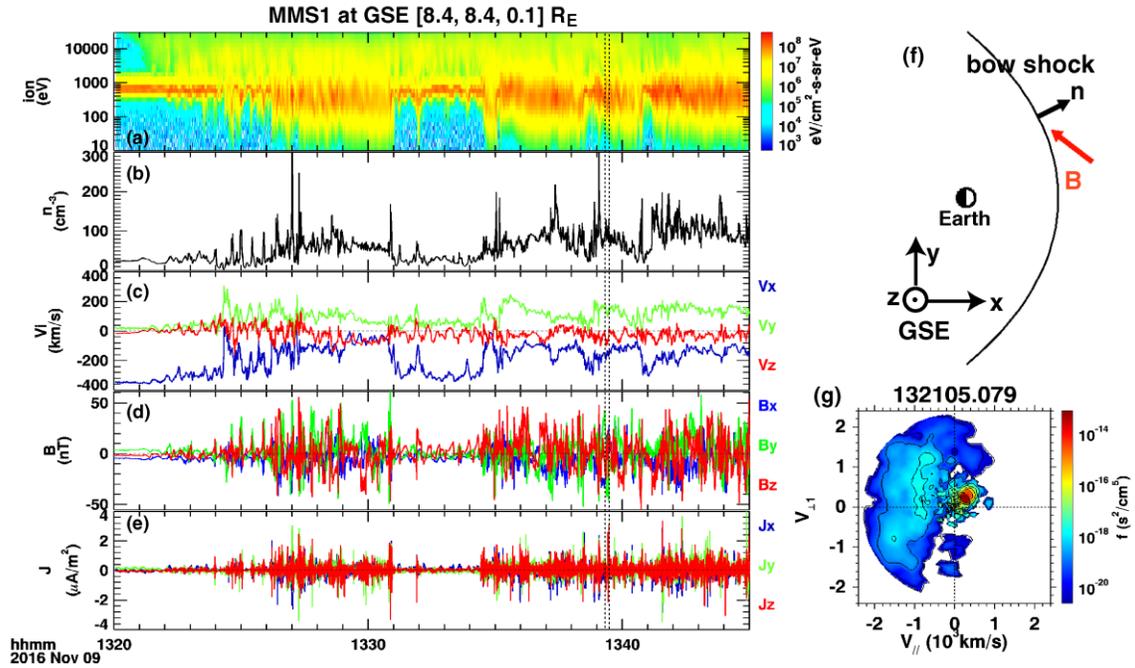

Figure 1. Overview of the shock observed by MMS1. The crossing is from the solar wind with cold, tenuous and fast ions to the magnetosheath with hot, dense and slow ions, seen from the ion spectrogram (a), electron density (b), and the ion bulk velocity (in GSE) (c). Significant magnetic field fluctuations exist (d), corresponding to current layers (e). (f) illustration of the observation location and upstream magnetic field orientation at the observing point. The current sheet between the two vertical lines is further analyzed in later figures. The much higher density than that in upstream and the decreasing $V_{ix}$ suggest the marked interval to be in the transition region before entering the magnetosheath proper. (g) Example ion distribution in the $\hat{V}_{\parallel} - \hat{V}_{\perp 1}$ plane upon entering the foreshock region, where $\hat{V}_{\perp 1} = (\hat{b} \times \hat{V}) \times \hat{b}$ and $\hat{V}$ is along the bulk velocity. $V_{\parallel}$ of suprathermal ions (blue) are negative, opposite to that of incoming solar wind ions (red), indicating that suprathermal ions are moving towards upstream.

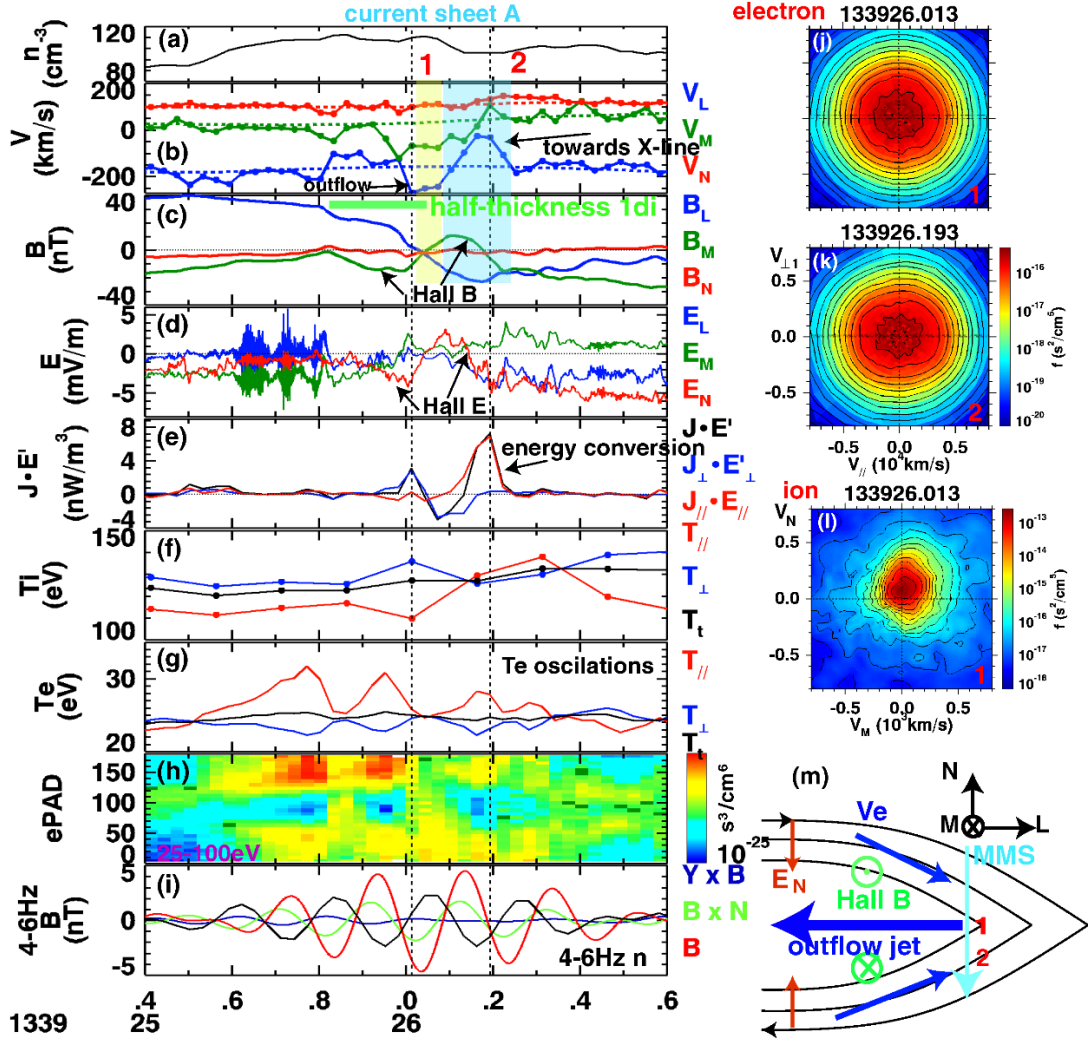

Figure 2. Current sheet A with features consistent with a reconnection diffusion region. (a) electron density is roughly symmetric on two sides of the current sheet; (b) electron velocity (solid) exhibits a flow towards the X-line sandwiching the outflow jet, while the ion velocity (dashed) remains the same with ambient values; (c) magnetic fields showing bipolar Hall $B_M$; the yellow and light blue shaded regions mark the intervals with positive and negative $V_{eL}$-$B_L$ correlations, respectively, as a way to identify reconnection outflows. (d) electric fields showing bipolar Hall $E_N$. LMN coordinate: L=[0.3409, -0.5671, 0.7497], M=[0.0087, 0.7994, 0.6007], N=[-0.9400, -0.1983, 0.2775] GSE. (e) energy conversion J · E′ has enhancements, where the electric field data are averaged over 30 ms to match the electron measurement cadence. (f) ion temperature. (g) electron temperature with oscillations in $T_{//}$ and $T_\perp$, while $T_t$ remains constant. (h) electron PA distributions: at $T_{//}$ peaks, field-aligned phase-space densities (psd) are enhanced; at $T_\perp$ peaks, PA~90° psd is enhanced and distributions are quasi-isotropic. (i) magnetic field fluctuations in the same frequency range with $T_e$, where the fluctuation along the background magnetic field (red) is anti-correlated with the density fluctuation (black). (j)-(k) electron reduced VDFs in the $\hat{V}_{||} - \hat{V}_{\perp 1}$ plane at locations 1 and 2. (l) ion VDF in the $V_M$-$V_N$ plane at location 1. (m) illustration of the current sheet crossing.

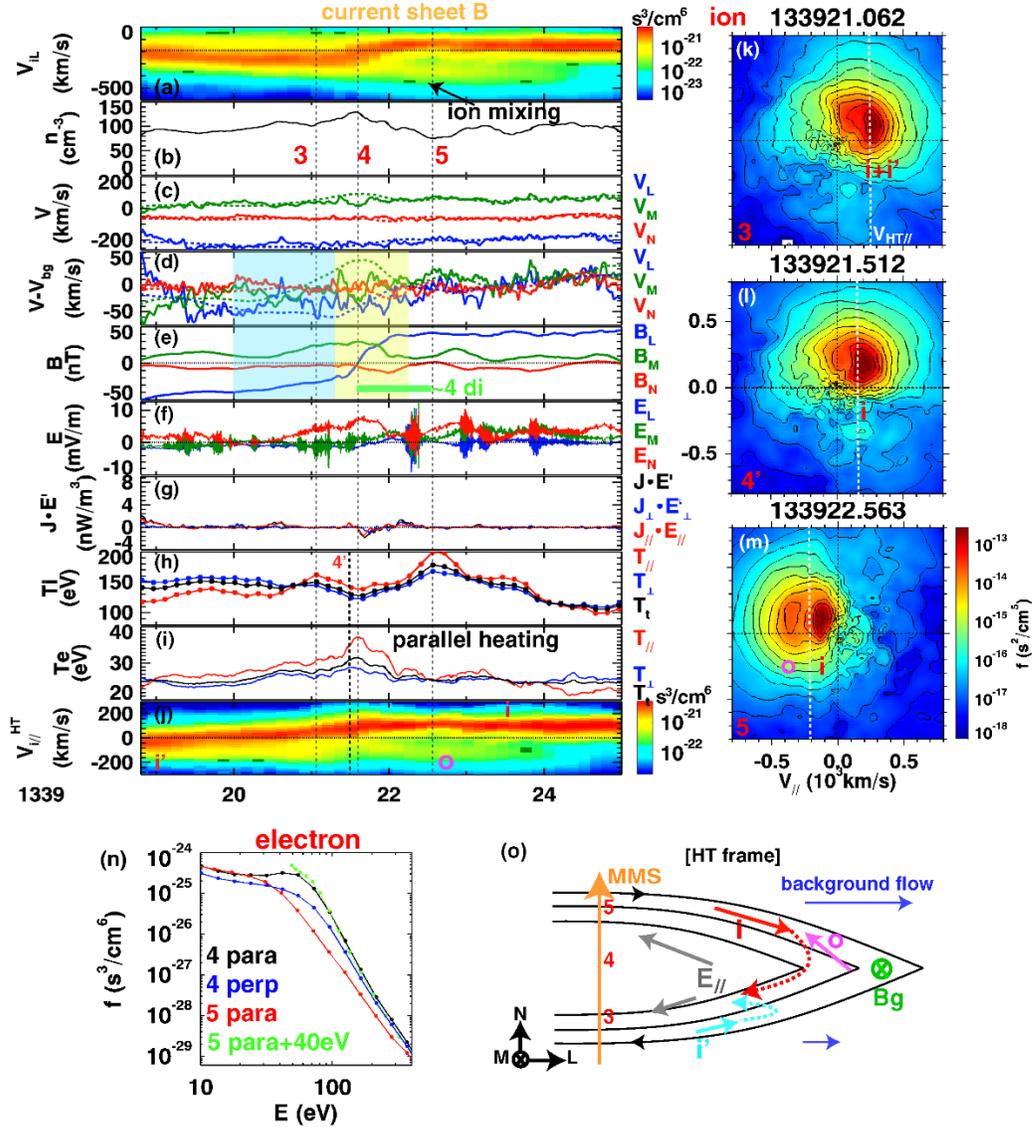

Figure 3. (a) ion spectrogram along $v_L$ cut at the bulk $V_M$ and $V_N$, showing mixing of multiple populations. The horizontal line marks the average $V_L$ between two sides of the inflow regions, taken as the $L$ component of the background flow. (b)-(i) same formats as in Figures 2a-2i, except that (d) shows the velocity after subtracting the background flow. LMN coordinate: L=[0.4725, -0.8527, 0.2227], M=[-0.3155, -0.3997, -0.8607], N=[0.8229, 0.3364, -0.4579] GSE. Ions and electrons both exhibit parallel heating. (j) ion $v_{//}$ spectrogram in the HT frame, cut at the bulk $V_\perp$, showing asymmetry between $B_L<0$ and $B_L>0$. Letters $i$, $i'$, and $o$ represent the two inflow and populations and the outflow component, respectively. (k)-(m) ion VDFs at locations 3-5, where the white dashed line marks the parallel component of the HT velocity. The two field-aligned populations in VDF 5 are interpreted as inflowing and outflowing ions as illustrated in (m), suggesting a reconnection structure. (n) 1D electron VDF along $v_{//}>0$ at location 4 (black) and 5 (red), and along $v_\perp$ at location 4 (blue). The green dashed curve is the VDF along $v_{//}>0$ at location 5 shifted by 40 eV, which matches VDF 4. The inflowing ion population is decelerated towards the mid-plane as inflowing electrons are accelerated. (o) illustration in the HT frame.

Geophysical Research Letters
Supporting information for

# Evidence of magnetic reconnection in the terrestrial bow shock transition region

Shan Wang[1,2], Li-Jen Chen[2], Naoki Bessho[1,2], Michael Hesse[3,4], Lynn B. Wilson III[2], Barbara Giles[2], Thomas E. Moore[2], Christopher T. Russell[5], Roy B. Torbert[6], and James L. Burch[4]

[1]Astronomy Department, University of Maryland, College Park, MD 20742
[2]NASA Goddard Space Flight Center, Greenbelt, MD 20771
[3]Department of Physics and Technology, University of Bergen, Bergen, Norway
[4]Southwest Research Institute San Antonio, San Antonio, TX 78238
[5]Department of Earth, Planetary, and Space Sciences, University of California, Los Angeles, CA 90095
[6]Space Science Center, University of New Hampshire, Durham, NH 03824
*swang90@umd.edu

**Contents of this file**
Text S1
Figure S1

**Introduction**
This supplementary information provides details about the analyses for electromagnetic waves that modulate the electron temperature in the observed reconnection diffusion region (current sheet A shown in Figure 2).

**Text S1**
For the electromagnetic fluctuations observed in the diffusion region, we determine that the kinetic Alfvén wave (KAW) is a candidate wave mode. The spacecraft-frame phase velocity ($V_{ph,sc}$) is determined by the correlation analysis of $\delta B_\parallel$ measured by four spacecraft during three intervals represented by the lengths of the blue lines in Figure S1b (N component of $V_{ph,sc}$). The three intervals are at $B_L>0$, near $B_L=0$, and at $B_L<0$, respectively, and they are individually calculated because $V_{iN}$ has a clear variation (black in Figure S1b). $V_{ph,sc}$ in LMN for the three intervals are: 95×[0.036, -0.103, 0.994] km/s, 102×[0.023, -0.100, 0.995] km/s, and 118×[0.001, -0.097, 0.995] km/s. Thus, the propagation is mainly along N with a small component along -M. According to the Doppler shift relation $\omega_{sc} = \omega + \boldsymbol{k} \cdot \boldsymbol{V_i}$, the ion-frame phase speeds for the three intervals are similarly to be 4±1 km/s, 7±4 km/s, and 8±5 km/s, where the uncertainty is based on the variation of $\boldsymbol{V_i}$ during each interval. Combining the results from the three intervals, the ion-frame frequency is 0.23-0.77 $f_{ci}$, where $f_{ci}$ is the ion cyclotron frequency right outside of the current sheet (0.7 Hz), and the wave number satisfies $kr_{gi}$=8.5±0.8. Taking $V_{iN}$ at $B_L$=0 of 103 km/s to be the current sheet motion, the increase of $V_{iN}$ is consistent with the inflow towards the current sheet mid-plane. The N component of $\boldsymbol{V_{ph}}$ in the current sheet frame also changes from negative to positive, indicating that the wave front propagates towards the current sheet mid-plane.

The theoretical parallel phase speed for KAW is $V_{ph\parallel} = V_A\sqrt{1 + k_\perp^2(r_{gi}^2 + r_{si}^2) - \frac{f^2}{f_{ci}^2}(1 + k_\perp^2 r_{gi}^2)}$, where $r_{si}$ is the ion gyro-radius based on the sound speed $\sqrt{k_B T_e/m_i}$ (Stasiewicz et al., 2000). The angle between $\mathbf{k}$ and $\mathbf{B_0}$ is calculated from the Minimum Variance Analysis of magnetic field fluctuations to be 88.9°. Using the average $|\mathbf{k}|$=0.3 km$^{-1}$ as $k_\perp$ ($k_\perp \gg k_{//}$) and parameters at the start of the wave fluctuations at 13:39:25.5 UT ($V_A$=80 km/s), the ion-frame $V_{ph//}$ is determined to be $5.2 \pm 3.2 V_A$, close to the predicted $7.8 \pm 1.3 V_A$. Together with the below-$f_{ci}$ frequency and the anti-correlated $\delta B_\parallel$-$\delta n$, the wave properties are consistent with KAWs.

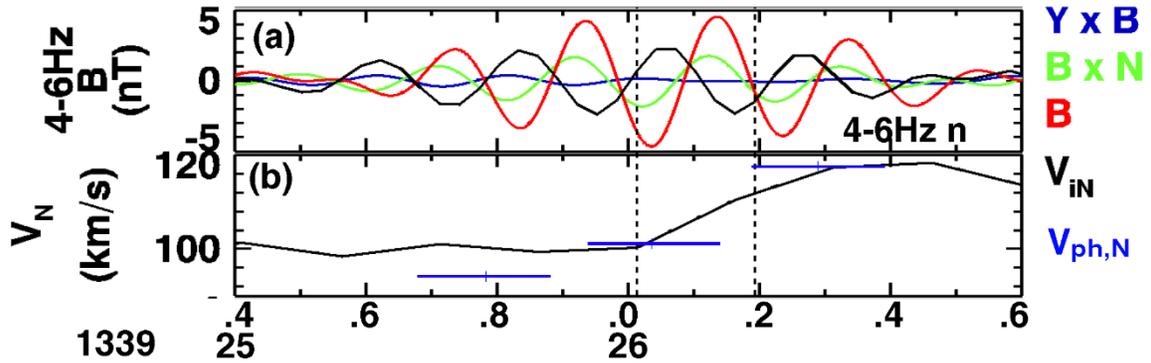

**Figure S1.** Properties of waves observed to modulate electron temperature in the reconnection diffusion region. (a) fluctuations of magnetic fields and electron densities. (b) Ion velocity normal to the current sheet (black), and the normal component of the wave phase velocity (blue).